\documentclass[12pt,a4paper]{article}
\usepackage{amsmath,amssymb}
\usepackage{epsfig}
\begin{document}

\begin{center}
\Large\textbf{ REGULAR INFLATIONARY COSMOLOGY AND GAUGE THEORIES OF GRAVITATION}\\
\normalsize A. V. Minkevich\\
 \textit{\small $^1$Department of Theoretical Physics, Belarussian State University,\\
av. F. Skoriny 4, 220050, Minsk, Belarus, phone: +(375)(17)2095114, fax: +(375)(17)2095445\\
 $^2$Department of Physics and Computer Methods, Warmia and Mazury University in Olsztyn,
 Poland\\
\normalsize  E-mail: MinkAV@bsu.by; awm@matman.uwm.edu.pl}
\end{center}
\begin{flushright}
\begin{minipage}{\textwidth}\small
\textbf{Abstract.} Cosmological equations for homogeneous
isotropic models filled by scalar fields and ultrarelativistic matter are
investigated in the framework of gauge theories of gravity. Regular inflationary
cosmological models of flat, closed and open type with dominating ultrarelativistic
matter at a bounce are discussed. It is shown that  essential part of
inflationary cosmological models has bouncing character.
\end{minipage}
\end{flushright}

\section{Introduction}
It is known that inflationary scenario plays the important role in
the early Universe theory and allows to resolve a number of
problems of standard Friedmann cosmology \cite{m1,Kolb}. Most
inflationary models built in the frame of general relativity
theory (GR) are singular and limited in the time in the past. In
the case of flat and open models such situation is inevitable
\cite{mn1}, and the quantum gravity era cannot be avoided in the
past\footnote{Recently the Pre-Big Bang Scenario in String
Cosmology was discussed (See \cite{Gasper} and Refs. given
therein).}. At the same time regular inflationary bouncing
solutions exist in GR for closed models filled by massive or
nonlinear scalar fields and usual matter [5-12]\footnote{Eternally
inflating regular closed model of an Emergent Universe with
asymptotically Einstein static state was discussed in Ref.
\cite{elis}.}. Because limiting (maximum) energy density and
limiting temperature at a bounce in such models can be essentially
less than the Planckian ones, classical description of
gravitational field in these models is possible and quantum
gravitational effects are negligible. Note that regular
inflationary solutions for closed models in GR are unstable at
compression stage, and small variations of gravitational and
scalar fields at compression lead to singularity because of the
divergence of the time derivative of scalar field $\dot{\phi}$. As
result the energy density
$\rho_{\phi}=\frac{1}{2}\dot{\phi}^2+V(\phi)$ and the pressure
$p=\frac{1}{2}\dot{\phi}^2-V(\phi)$  ($V$ is a scalar field
potential) of scalar field at compression stage near singularity
are connected in the following way $p_\phi\approx\rho_\phi$
\cite{m6}. In the case of regular bouncing solutions the scalar
field potential has to be dominating at a bounce, so in the case
of models including scalar field $\phi$ and ultrarelativistic
matter with the energy density $\rho_r$ the following condition at
a bounce must be valid
\begin{equation}
V-\dot{\phi}^2-\rho_{r}>0.
\end{equation}

The relation (1) means that the greatest part of energy density at
the end of cosmological compression has to be determined by scalar
fields\footnote{Note that classical solutions of cosmological Friedmann
equations of GR for models including scalar fields can be used
only for limited time intervals because of the processes of mutual
transformations of elementary particles and scalar fields in the
early Universe.}.  Because there are not physical reasons ensuring
the realization of condition (1), regular inflationary solutions
in GR cannot be considered as a base to build regular inflationary
cosmology.

Different situation takes place in gauge theories of gravitation (GTG) - Poincare
GTG, metric-affine GTG. Note that GTG are natural generalization of GR by applying
the local gauge invariance principle to gravitational interaction (see review
\cite{Hehl}). As it was shown in a number of our papers [15-20,12], the GTG possess
important regularizing properties. By satisfying the correspondence principle with
GR in the case of usual gravitating systems with rather small energy densities and
pressures, GTG can lead to essentially different physical conclusions in the case of
gravitating systems at extreme conditions with extremely high energy densities and
pressures \cite{m10a,m2b}. In particular, gravitating vacuum with sufficiently high
energy density $\rho_v>0$    and pressure  $p_v=-\rho_v$  can lead to the vacuum
gravitational repulsion effect and to a bounce in closed, open and flat models, that
allows to build regular inflationary cosmological models
\cite{m2,m7a,mink21b}. Some particular regular inflationary cosmological
solutions for closed models with scalar fields were discussed in
Refs.\cite{m7b,mink21a,mink21b}. The present paper is devoted to study the problem
what place take regular inflationary cosmological models in GTG.

\section{Generalized cosmological Friedmann equations in GTG}

Homogeneous isotropic models in GTG are described by the following generalized cosmological
Friedmann equations (GCFE)
\begin{equation}
\frac{k}{R^2}+\left\{\frac{d}{dt}\ln\left[R\sqrt{\left|1-\beta\left(\rho-
3p\right)\right|}\,\right]\right\}^2=\frac{8\pi}{3M_p^2}\,\frac{\rho-
\frac{\beta}{4}\left(\rho-3p\right)^2}{1-\beta\left(\rho-3p\right)}
\, ,
\end{equation}
\begin{equation}
\frac{\left[\dot{R}+R\left(\ln\sqrt{\left|1-\beta\left(\rho-
3p\right)\right|}\,\right)^{\cdot}\right]^\cdot}{R}=
-\frac{4\pi}{3M_p^2}\,\frac{\rho+3p+\frac{\beta}{2}\left(\rho-3p\right)^2}{
1-\beta\left(\rho-3p\right)}\, ,
\end{equation}
where $R(t)$ is the scale factor of Robertson-Walker metrics, $k=+1,0,-1$ for
closed, flat, open models respectively, $\rho$  is energy density, $p$  is pressure,
$\beta$  is indefinite parameter with inverse dimension of energy density, $M_p$  is
Planckian mass. (The system of units with $\hbar=c=1$ is used). At first the GCFE
were deduced in Poincare GTG \cite{m10a}, and later it was shown that Eqs.(2)-(3)
take place also in metric-affine GTG \cite{m11a,m12a}. From Eqs. (2)-(3) follows the
conservation law in usual form
\begin{equation}
\dot{\rho}+3H\left(\rho+p\right)=0,
\end{equation}
where $H=\frac{\dot{R}}{R}$   is the Hubble parameter. Besides cosmological equations (2)-(3) gravitational
equations of GTG lead to the following relation for torsion
function $S$ and nonmetricity function $Q$
\begin{equation}
S-\frac{1}{4}Q=-\frac{1}{4}\,\frac{d}{dt}
\ln\left|1-\beta(\rho-3p)\right|.
\end{equation}
In Poincare GTG $Q=0$ and Eq. (5) determines the torsion function. In metric-affine
GTG there are three kinds of models \cite{m12a}: in the Riemann-Cartan space-time
($Q=0$), in the Weyl space-time ($S=0$), in the Weyl-Cartan space-time ($S\neq 0$,
$Q\neq 0$, the function S is proportional to the function $Q$). The value of
$|\beta|^{-1}$ determines the scale of extremely high energy densities. The GCFE
(2)-(3) coincide practically with Friedmann cosmological equations of GR if the
energy density is small $\left|\beta(\rho-3p)\right|\ll 1$. The difference between
GR and GTG can be essential at extremely high energy densities
$\left|\beta(\rho-3p)\right|\gtrsim 1$.\footnote{Ultrarelativistic matter with
equation of state $p=\frac{1}{3}\rho$ is exceptional system because Eqs. (2)--(3)
are identical to Friedmann cosmological equations of GR in this case independently
on values of energy density.} In the case of gravitating vacuum with constant energy
density $\rho_v=\mathrm{const}>0$ the GCFE (2)--(3) are reduced to Friedmann
cosmological equations of GR and $S=Q=0$, this means that de Sitter solutions for
metrics with vanishing torsion and nonmetricity are exact solutions of GTG
\cite{m13a,m14a} and hence inflationary models can be built in the frame of GTG.

In order to analyze inflationary cosmological models in GTG let us
consider systems including scalar field   minimally coupled with gravitation
and usual matter in the form of ultrarelativistic matter.
This assumption is available by analysis of the hot Universe models in the
beginning of cosmological expansion. In the case of other form of gravitating matter
our consideration would be more complicated.
If the interaction between scalar field and ultrarelativistic matter is negligible, the energy density
$\rho$  and pressure $p$   take the form
\begin{equation}
\rho=\frac{1}{2}\dot{\phi}^2+V(\phi)+\rho_r, \qquad
p=\frac{1}{2}\dot{\phi}^2-V(\phi)+\frac{1}{3}\rho_r,
\end{equation}
and the conservation law (4) leads to the scalar field equation
\begin{equation}
\ddot{\phi}+3H\dot{\phi}=-V' \qquad
\left(V'=\frac{dV}{d\phi}\right)
\end{equation}
and the conservation law for matter, which in our case has the
following integral ${\rho_rR^4={\rm const}}$. By using Eqs.
(6)-(7) the GCFE (2)-(3) can be written in the following form
\begin{eqnarray}
& & \frac{k}{R^2}
Z^2+\left\{H\left[1-2\beta(2V+\dot{\phi}^2)\right]-3\beta
V'\dot{\phi}\right\}^2\nonumber \\
& & \phantom{+H\left[1-2\beta(2V+\dot{\phi}^2)\right]} =\frac{8
\pi}{3M_p^2}\,\left[\rho_r+
\frac{1}{2}\dot{\phi}^2+V-\frac{1}{4}\beta\left(4V-\dot{\phi}^2
\right)^2\right]Z,\\
& &\dot{H}\left[1-2\beta(2V+\dot{\phi}^2)\right]Z+H^2\left\{\left[
1-4\beta(V-4\dot{\phi}^2)\right]Z-18\beta^2\dot{\phi}^4\right\}\nonumber\\
& &\phantom{H} +12\beta
H\dot{\phi}V'\left[1-2\beta(2V+\dot{\phi}^2)\right]-
3\beta\left[(V''\dot{\phi}^2-V'{}^2)Z+6\beta\dot{\phi}^2
V'{}^2\right]
\nonumber\\
& &\phantom{\dot{H}\left[1-2\beta(2V+\dot{\phi}^2)\right]Z}
=\frac{8 \pi}{3M_p^2}\,\left[V-\dot{\phi}^2-\rho_r
-\frac{1}{4}\beta(4V-\dot{\phi}^2)^2\right]Z,
\end{eqnarray}
where $Z=1-\beta(4V-\dot{\phi}^2)$, $\displaystyle
V'=\frac{dV}{d\phi}$, $\displaystyle V''=\frac{d^2V}{d\phi^2}$.
Relation (5) takes the form
\begin{equation}
S-\frac{1}{4}Q=\frac{3\beta}{2}\,\frac{\left(H\dot{\phi}+V'\right)\dot{\phi}}{1-\beta
\left(4V-\dot{\phi}^2\right)}.
\end{equation}
Unlike GR the cosmological equation (8) leads to essential
restrictions on admissible values of scalar field and permits to
exclude the divergence of derivative $\dot{\phi}$ for any finite
value of $\phi$, if $\beta<0$. Imposing $\beta<0$, we obtain from
Eq. (8) in the case $k=0,+1$
\begin{equation}
Z\geq 0 \qquad \text{or}\qquad \dot{\phi}^2\leq 4V+|\beta|^{-1}.
\end{equation}
Inequality (11) is valid also for open models discussed in Sec. 3. The region $\Sigma$ of admissible
values of scalar field on the plane $P$ with the axis  ($\phi$, $\dot{\phi}$) determined by (11) is limited by two bounds
$L_{\pm}$
\begin{equation}
\dot{\phi}=\pm \left(4V+|\beta|^{-1}\right)^{\frac{1}{2}}.
\end{equation}
From Eq. (8) the Hubble parameter on the bounds $L_{\pm}$  is equal to
\begin{equation}
  H=\frac{3\beta V'\dot{\phi}}{1-2 \beta(2V+\dot{\phi}^2)}.
\end{equation}
According to (13) the right-hand part of Eq. (10) is equal to
$\frac{1}{2}H$ on the bounds $L_{\pm}$. This means that the
torsion (nonmetricity) will be regular, if the Hubble parameter
and scalar field are regular. In connection with this our main
attention in Sec. 3 will be turned to study properties of
solutions of GCFE (8)-(9). In accordance with (11) energy density
$\rho_{\phi}$  and pressure $p_{\phi}$ of scalar field satisfy the
condition $p_{\phi}\leq \frac{1}{3}(\rho_{\phi}+|\beta|^{-1})$.
Unlike GR the equation of state for scalar field
$p_{\phi}\approx\rho_{\phi}$   is not valid at any stage of models
evolution in GTG.

\section{Regular inflationary models in GTG}

Let us consider the most important general properties of cosmological solutions of
GCFE (8)-(9). At first, note by given initial conditions for scalar field
($\phi$,$\dot{\phi}$) and values of $R$ and $\rho_r$ there are two different
solutions corresponding to two values of the Hubble parameter following from
Eq.~(8):
\[ 
  H_{\pm}=\frac{3\beta V'\dot{\phi}\pm\sqrt{D}}{1-2 \beta(2V+\dot{\phi}^2)},
\] 
where
\[
D=\frac{8\pi}{3M_p^2}\,\left[\rho_r+
\frac{1}{2}\dot{\phi}^2+V-\frac{1}{4}\beta\left(4V-\dot{\phi}^2
\right)^2\right]Z-\frac{k}{R^2}\,Z^2\ge 0
\]
Unlike GR, the values of $H_{+}$ and $H_{-}$ in GTG are sign-variable and, hence,
both solutions corresponding to $H_{+}$ and $H_{-}$ can describe the expansion as
well as the compression in dependence on their sign. Below we will call solutions of
GCFE corresponding to $H_{+}$ and $H_{-}$ as $H_{+}$-solutions and $H_{-}$-solutions
respectively. In points of bounds $L_{\pm}$ we have $D=0$, $H_{+}=H_{-}$ and the
Hubble parameter is determined by (13). The bounds $L_{\pm}$ are particular curves
for GCFE. If $Z=0$, Eqs. (8)--(9) are satisfied, and because the bounds $L_\pm$ are
not limited for applying potentials $V(\phi)$ and tend to infinity under
$|\phi|\to\infty$, corresponding solutions of GCFE are singular.

In order to study the behaviour of cosmological models at the beginning of
cosmological expansion, let us analyze extreme points for the scale factor $R(t)$:
$R_0=R(0)$, $H_0=H(0)=0$. Denoting values of quantities at $t=0$ by means of index
"0", we obtain  from (8)-(9):
\begin{eqnarray}
& &\frac{k}{R_0^2} Z_0^2+9\beta^2 V'{}_0^2\dot{\phi}_0^2=\frac{8
\pi}{3M_p^2}\,\left[\rho_{r0}+\frac{1}{2}\dot{\phi_0}^2
+V_0-\frac{1}{4}\beta\left(4V_0-\dot{\phi}_0^2\right)^2\right]Z_0,\\
& &\dot{H}_0=\left\{\frac{8
\pi}{3M_p^2}\,\left[V_0-\dot{\phi}_0^2-\rho_{r0}-\frac{1}{4}\beta(4V_0
-\dot{\phi}_0^2)^2\right]Z_0\right.\nonumber\\
&
&\left.\phantom{H_0=}+3\beta\left[(V''{}_0\dot{\phi}_0^2-V'{}_0^2)Z_0
+6\beta\dot{\phi}_0^2 V'{}_0^2\right]\right\}
\left[1-2\beta(2V_0+\dot{\phi}_0^2)\right]^{-1}Z_0^{-1},
\end{eqnarray}
where $Z_0=1-\beta(4V_0-\dot{\phi}_0^2)$. A bounce point is described by Eq. (14),
if the value of $\dot{H}_0$ is positive. By giving concrete form of potential $V$
and choosing values of $R_0$, $\phi_0$, $\dot{\phi}_0$ and $\rho_{r0}$ at a bounce,
we can obtain numerically particular bouncing solutions of GCFE for various values
of parameter $\beta$.

The analysis of GCFE shows, that the properties of cosmological solutions depend
essentially on the parameter  $\beta$, i.e. on the scale of extremely high energy
densities. From physical point of view interesting results can be obtained, if the
value of $|\beta|^{-1}$   is much less than the Planckian energy density
\cite{mink21b}, i.e. in the case of large in module values of parameter $\beta$ (by
imposing $M_p=1$). In order to investigate cosmological solutions at the beginning
of cosmological expansion in this case, let us consider the GCFE by supposing that
\begin{equation}
\left|\beta\left(4V-\dot{\phi}^2\right)\right|\gg 1,\qquad
\rho_r+\frac{1}{2}\dot{\phi}^2+V\ll\left|\beta\right|
\left(4V-\dot{\phi}^2\right)^2.
\end{equation}
Note that the second condition (16) does not exclude that ultrarelativistic matter
energy density can dominate at a bounce $\rho_r\gg V+\frac{1}{2}\dot{\phi}^2$
\cite{mink21b}. We obtain:
\begin{equation}
\frac{k}{R^2}+\frac{\left[2H\left(2V+
\dot{\phi}^2\right)+3V'\dot{\phi}\right]^2}{\left(4V-\dot{\phi}^2\right)^2}
=\frac{2\pi}{3M_p^2} \left(4V-\dot{\phi}^2\right),
\end{equation}
\begin{multline}
  \dot{H}\left(2V+\dot{\phi}^2\right)\left(4V-\dot{\phi}^2\right)+
  H^2\left(8V^2-34V\dot{\phi}^2-\dot{\phi}^4\right)-12H
  V'\dot{\phi}\left(2V+\dot{\phi}^2\right)\\
  + \frac{3}{2}\left(V''\dot{\phi}^2-V'^2\right)\left(4V-\dot{\phi}^2\right)-
  9V'^2\dot{\phi}^2
  =  \frac{\pi}{3M_p^2}\left(4V-\dot{\phi}^2\right)^3.
\end{multline}
Eqs. (17)-(18) do not include radiation energy density, which does not have influence on the dynamics
of inflationary models in the case under consideration (although, as it was noted above, the contribution
 of ultrarelativistic matter to energy density
can be essentially greater in comparison with scalar field), moreover Eqs. (17)-(18)
do not contain the parameter $\beta$. According to Eq. (17) the Hubble parameter in
considered approximation is equal to
\[
H_\pm=\frac{\displaystyle -3V'\dot{\phi}\pm \left|4V-\dot{\phi}^2\right|\,
\sqrt{\frac{2\pi}{3M_p^2}\left(4V-\dot{\phi}^2\right)-\frac{k}{R^2}}}{\displaystyle
2\left(2V+\dot{\phi}^2\right)}\, ,
\]
and  extreme points of the scale factor are determined by the following condition
\begin{equation}
\frac{k}{R_0^2}+9\left(\frac{V_0'\dot{\phi}_0}{4V_0-\dot{\phi}^2_0}\right)^2=
\frac{2\pi}{3M_p^2}\left(4V_0-\dot{\phi}^2_0\right).
\end{equation}
From Eq. (18) the time derivative of the Hubble parameter at extreme points is
\begin{equation}
\dot{H}_0=\left[\frac{\pi}{3M_p^2}\,\left(4V_0-\dot{\phi}_0^2\right)^{2}+
\frac{3}{2}\left(V_0'^2-V_0''\dot{\phi}_0^2\right)
+9V_0'^2\dot{\phi}_0^2\left(4V_0-\dot{\phi}_0^2\right)^{-1}\right]
\left(2V_0+\dot{\phi}_0^2\right)^{-1}.
\end{equation}
Obviously Eq. (19)-(20) correspond to (14)-(15) in considered approximation.

Now let us consider flat models, for which Eq. (19) can be written
in the following form
\begin{equation}
4V_0-\dot{\phi}_0^2=3\left(\frac{M_p^2}{2\pi}\,{V_0'}^{2}\,\dot{\phi}_0^2\right)^{\frac{1}{3}}.
\end{equation}
Solutions of Eq. (21) are in physical region $\Sigma$  on the plane $P$ determined
in considered approximation according to (11) by the following inequality
$4V-\dot{\phi}^2>0$  (the neighbourhood of origin of coordinates on the plane $P$ is
not considered in our approximation). Eq.(21) has as solutions two curves $B_1$ and
$B_2$ on the plane $P$,  which are situated near bounds $L_{+}$ and $L_{-}$
respectively. Because of (21) the bounce condition $\dot{H}_0>0$ can be written in
the form
\begin{equation}
\frac{2\pi}{3M_p^2}\,\left(4V_0-\dot{\phi}_0^2\right)^{2}+
V_0'^2-V_0''\dot{\phi}_0^2>0 \quad\text{or}\quad
6\left(\frac{\pi}{4M_p^2}\right)^{\frac{1}{3}}\left(V'_0\,\dot{\phi}_0\right)^{\frac{4}{3}}+
V_0'^2-V_0''\dot{\phi}_0^2>0.
\end{equation}
If the condition (22) is satisfied on the curves $B_{1,2}$, the bounce will take
place for corresponding solutions in points of these curves, which can be called
"bounce curves". Note that bounce curves exist in the case of various scalar field
potentials applying in inflationary cosmology, in particular, in the case of power
potentials $V=\alpha
\phi^{2n}$ ($\alpha=\mathrm{const}>0$ and $n$ is integer positive number). Each of
two curves $B_{1,2}$ contains two parts corresponding to vanishing of $H_{+}$ or
$H_{-}$ and denoting by ($B_{1+}$, $B_{2+}$) and ($B_{1-}$, $B_{2-}$) respectively.
If $V'$ is positive (negative) in quadrants 1 and 4 (2 and 3) on the plane $P$, the
bounce will take place in points of bounce curves $B_{1+}$ and $B_{2+}$ ($B_{1-}$
and $B_{2-}$) in quadrants 1 and 3 (2 and 4) for $H_{+}$-solutions
($H_{-}$-solutions) (see Fig. 1).
\begin{figure}[htb!]
\begin{minipage}{0.48\textwidth}\centering{
\epsfig{file=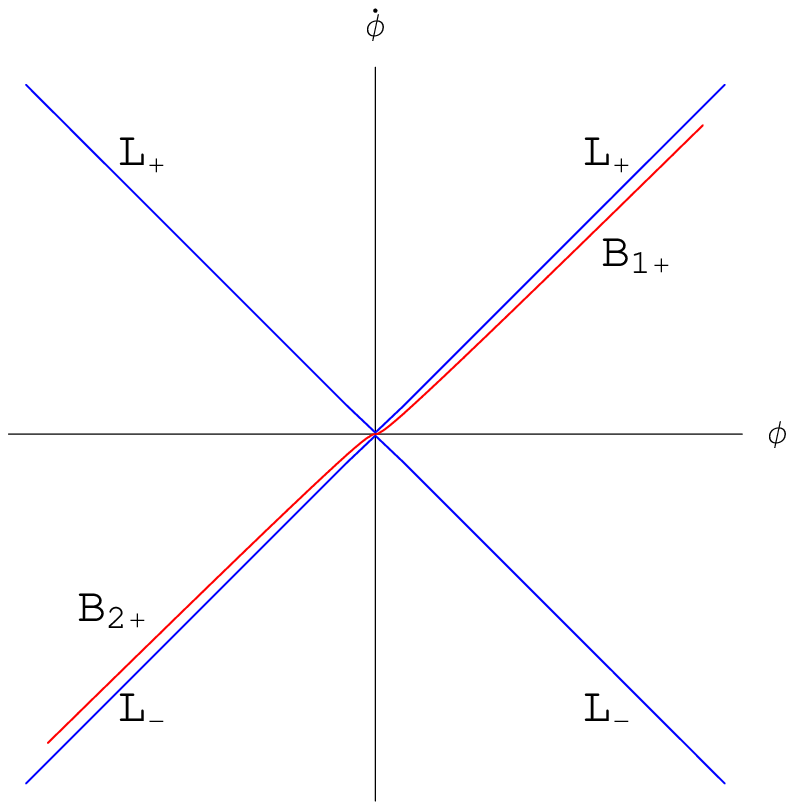,width=\linewidth}}
\end{minipage}\, \hfill\,
\begin{minipage}{0.48\textwidth}\centering{
\epsfig{file=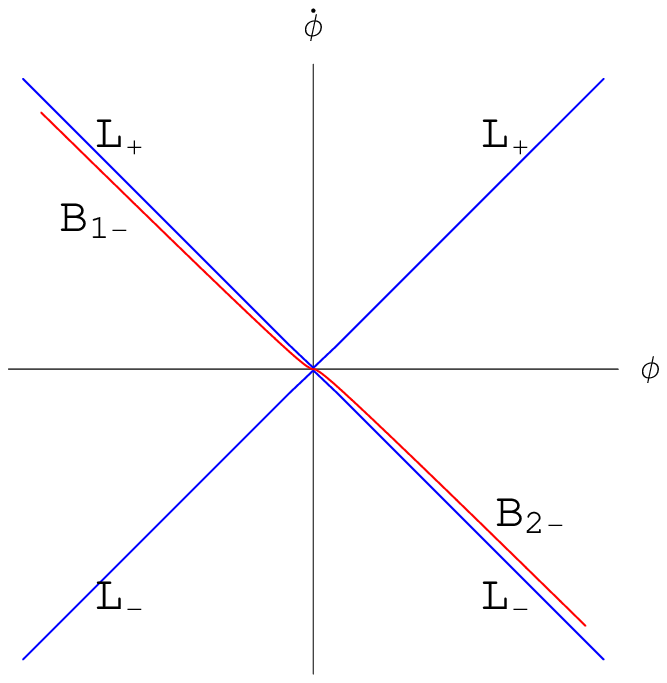,width=\linewidth}}
\end{minipage}
\caption[]{Bounce curves for $H_{+}$-solutions  and $H_{-}$-solutions in the case of
potential $V=\frac{1}{2}m^2\phi^2$.}
\end{figure}
As result practically all phase trajectories for $H_{+}$--solutions
($H_{-}$--solutions) reaching bounce curves $B_{1+}$ and $B_{2+}$ ($B_{1-}$ and
$B_{2-}$) correspond to bouncing solutions.  To obtain regular bouncing solutions we
have to take into account that bounds $L_{\pm}$ defined by relation $Z=0$ and
Eq.(13) correspond to particular solutions of GCFE ($L_{\pm}$--solutions), in points
of which $H_{-}$--solutions reach the bounds and $H_{+}$--solutions originate from
them\footnote{Remind that $H_{+}=H_{-}$ on the bounds $L_{\pm}$. The Hubble
parameter $H_{+}$ is negative in regions between curves ($L_{+}$ and $B_{1+}$),
($L_{-}$ and $B_{2+}$), and the value of $H_{-}$ is positive in regions between
curves ($L_{+}$ and $B_{1-}$), ($L_{-}$ and $B_{2-}$). In all other admissible
regions on the plane $P$ the sign of values $H_{+}$ and $H_{-}$ is normal:
$H_{+}>0$, $H_{-}<0$.}. As result the regular transition from $H_{-}$--solutions to
$H_{+}$--solutions is possible on bounds $L_{\pm}$, in particular, in the form of
their glueing. In consequence of this each solution containing as parts
$H_{-}$-solution and $H_{+}$-solution describes regular inflationary model, and a
bounce takes place by intersection of phase trajectory with corresponding bounce
curve. However, singular solutions exist also. As it was noted above in the case of
various potentials applying in inflationary cosmology (in particular, power
potentials) the curves $L_{\pm}$ are not limited on the plane P and tend to
infinity; the scalar field satisfying the following equation on bounds
$\ddot{\phi}=2V'$ diverges in the past and in the future, and hence particular
$L_{\pm}$ -solutions are singular. As result any solution containing
$H_{+}$-solution (or $H_{-}$-solution) glueed with particular $L_{\pm}$-solutions is
singular in the past (or in the future). Note the number of such singular solutions
is much less than the number of regular solutions, namely to one singular solution
correspond infinite number of regular solutions obtaining by glueing of noted above
singular solution with $H_{-}$-solution (or $H_{+}$-solution) in different points of
$L_{\pm}$-curve and excluding then divergent part of $L_{\pm}$ -solution.

 In the case of open and closed models Eq. (19) determines 1-parametric family of
bounce curves with parameter $R_0$. Bounce curves of closed models are situated on
the plane $P$ in region between two bounce curves $B_{1}$ and $B_{2}$ of flat
models, and in the case of open models bounce curves are situated in two regions
between the curves: $L_{+}$ and $B_{1}$, $L_{-}$ and $B_{2}$. Because the behaviour
of bounce curves for open models is like to that for flat models, the situation
concerning bouncing inflationary solutions in the case of open models is the same as
described above situation for flat models. Unlike flat and open models, for which
$H_{+}= H_{-}$ only in points of bounds $L_{\pm}$ and regular inflationary models
can be built if $H_{+}$-- and $H_{-}$--solutions reach bounds$L_{\pm}$, in the case
of closed models the regular transition from $H_{-}$-solution to $H_{+}$-solution is
possible without reaching the bounds $L_{\pm}$. It is because by certain value of
$R$ we have $H_{+}= H_{-}$ in the case if $Z\neq 0$. Similar regular inflationary
solution was considered in Ref.\cite{mink21b}. Note that bouncing solutions exist
not only in classical region, where scalar field potential and kinetic energy
density of scalar field go not exceed the Planckian energy density, but also in
regions, where classical restrictions on scalar fields are not fulfilled and
according to accepted opinion quantum gravitational effects can be essential.

In general case, when approximation (16) is not valid, bounce
curves of cosmological models determined by Eq.(14) depend on
parameter $\beta$ . By certain value of $\beta$ we have
1-parametric family of bounce curves with parameter $\rho_{r0}$
for flat models, and we have 2-parametric families of bounce
curves for closed and open models with parameters  $R_0$ and
$\rho_{r0}$. According to Eqs. (14)--(15) the bounce condition
$\dot{H}_0>0$ leads to the following relation
\begin{equation}
\frac{8
\pi}{3M_p^2}\,\left[V_0+\frac{1}{3}\rho_{r0}-\frac{1}{4}\beta\left(4V_0-\dot{\phi}_0^2\right)^2
\right]-\beta\left(V_0'^2-V_0''\dot{\phi}_0^2\right)-\frac{2k}{3R_0^2}\,
Z_0>0
\end{equation}
It is follows from (23),unlike GR the presence of ultrarelativistic matter does not
prevent from the bounce realization (compare with (1)) by reaching certain bounce
curve. We see the GCFE for homogeneous isotropic models including scalar fields and
ultrarelativistic matter allow to build regular inflationary cosmological models of
flat, open and closed type, although singular solutions exist also. The problem of
excluding singular solutions in inflationary cosmology in the frame of GTG is
analyzed in Ref.\cite{m14b}, and solutions of GTG near the origin of coordinates on
the plane $P$ are discussed in Ref.\cite{m14c}.

 As illustration of obtained results we
will consider particular bouncing cosmological inflationary solution for flat model
by using scalar field potential in the form $V=\frac{1}{2}m^2\phi^2$
(${m=10^{-6}M_p}$). The solution was obtained by numerical integration of Eqs. (7),
(9) and by choosing in accordance with Eq.(14) (or (21)) the following values of
scalar field at a bounce: $\phi_0=\sqrt{2}\, 10^3\, M_p$,
$\dot{\phi}_0=\sqrt{3.96757 V_0\,}$ ($\beta=-10^{14}M_p^{-4}$). A bouncing solution
includes: quasi-de-Sitter stage of compression, the stage of transition from
compression to expansion, quasi-de-Sitter inflationary stage, stage after inflation.
The dynamics of the Hubble parameter and scalar field is presented for different
stages of obtained bouncing solution in Figures 2--4 (by choosing $M_p=1$).
\begin{figure}[htb!]
\begin{minipage}{0.48\textwidth}\centering{
\epsfig{file=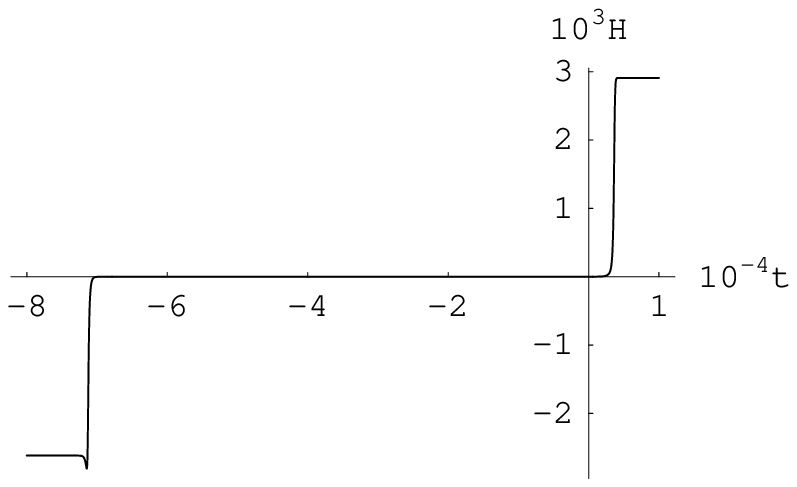,width=\linewidth}}
\end{minipage}\, \hfill\,
\begin{minipage}{0.48\textwidth}\centering{
\epsfig{file=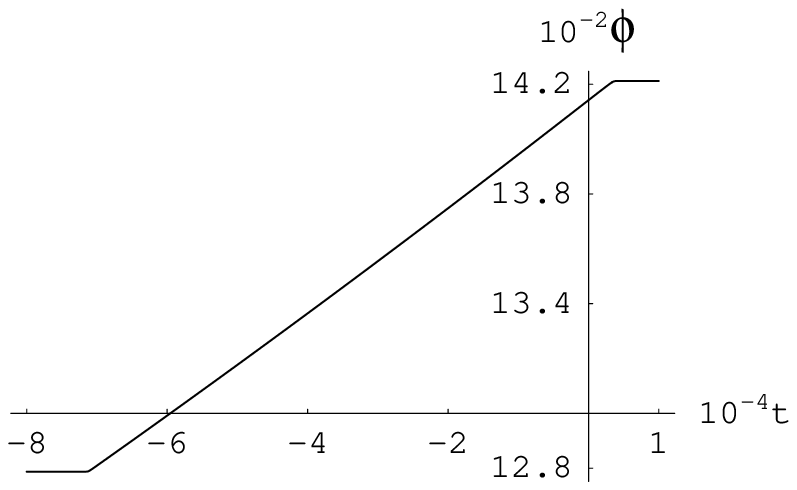,width=\linewidth}}
\end{minipage}
\caption{The stage of transition from compression to expansion.}
\end{figure}
\begin{figure}[htb!]
\begin{minipage}{0.48\textwidth}\centering{
\epsfig{file=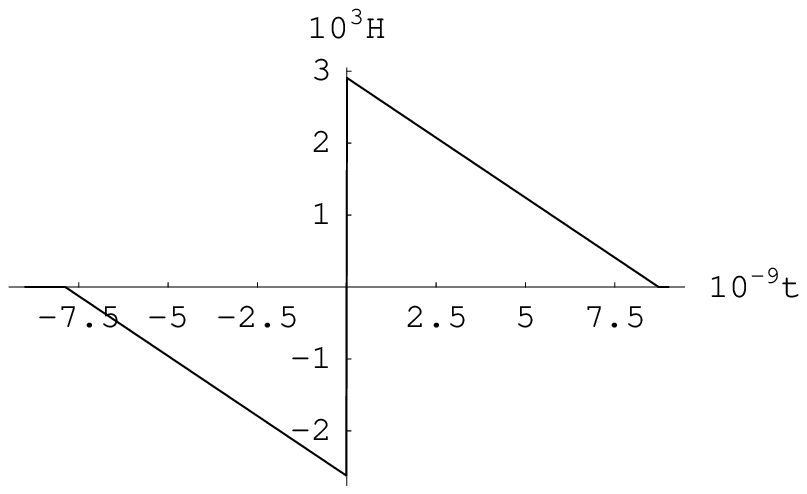,width=\linewidth}}
\end{minipage}\, \hfill\,
\begin{minipage}{0.48\textwidth}\centering{
\epsfig{file=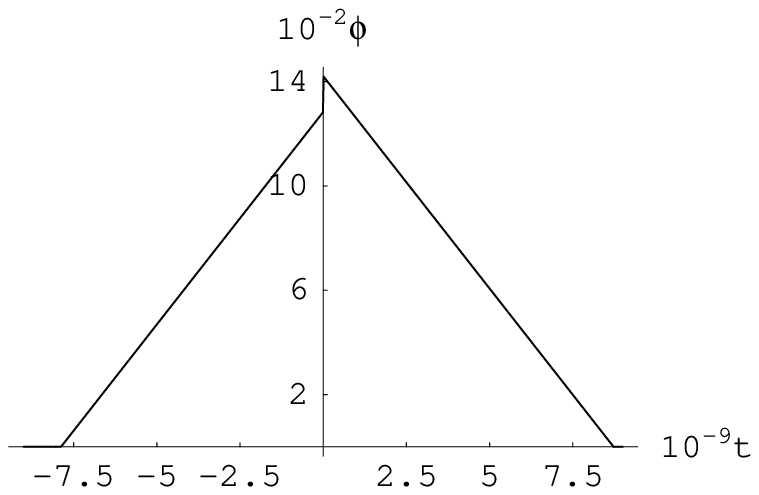,width=\linewidth}}
\end{minipage}
\caption{Quasi-de-Sitter stage of compression and inflationary stage.}
\end{figure}
\begin{figure}[htb!]
\begin{minipage}{0.48\textwidth}\centering{
\epsfig{file=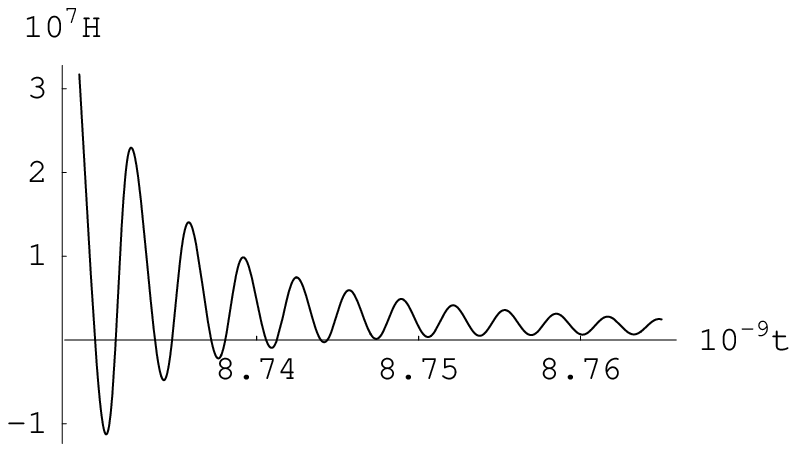,width=\linewidth}}
\end{minipage}\, \hfill\,
\begin{minipage}{0.48\textwidth}\centering{
\epsfig{file=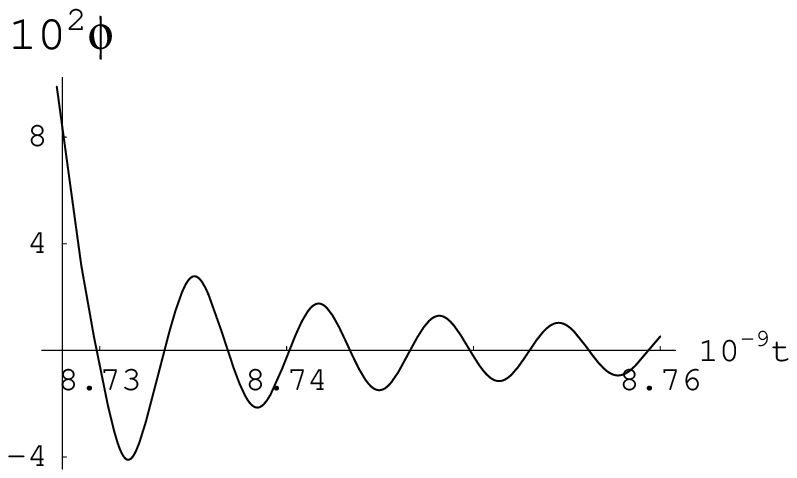,width=\linewidth}}
\end{minipage}
\caption{The stage after inflation.}
\end{figure}
The transition stage from compression to expansion (Fig. 2) is essentially
asymmetric with respect to the point $t=0$ because of $\dot{\phi}_0\neq 0$. Similar
asymmetry is inevitable property of bouncing solutions for flat and open models. In
course of transition stage the Hubble parameter changes from maximum in module
negative value at the end of compression stage to maximum positive value at the
beginning of expansion stage. The scalar field changes linearly at transition stage,
the derivative $\dot\phi$ grows at first from positive value  $\dot{\phi}_1\sim
1.6\cdot 10^{-7}$  to maximum value  $\dot{\phi}\sim\dot{\phi}_0$ and then decreases
to negative value $\dot{\phi}_2\sim -1.6\cdot 10^{-7}$. Quasi-de-Sitter inflationary
stage and quasi-de-Sitter compression stage are presented in Fig.3. Although the
GCFE (8)-(9) and their approximation (17)-(18) have different structure from
cosmological Friedmann equations of GR, like GR the time dependence of functions
$H(t)$ and $\phi(t)$  at compression  and inflationary stages is linear. The
amplitude and frequency of oscillating scalar field after inflation (Fig. 4) are
different than that of GR, this means that approximation of small energy densities
$\left|\beta\left(4V-\dot{\phi}^2\right)\right|\ll 1$ at the beginning of this stage
is not valid; however, the approximation (17)--(18) is not valid also because of
dependence on parameter $\beta$  of oscillations characteristics, namely, amplitude
and frequency of scalar field oscillations decrease by increasing of $|\beta|$
\cite{mink21b}. The behaviour of the Hubble parameter after inflation is also
noneinsteinian, at first the Hubble parameter oscillates near the value $H=0$, and
later the Hubble parameter becomes positive and decreases with the time like in GR.
Before quasi-de-Sitter compression stage there are also oscillations of the Hubble
parameter and scalar field not presented in Figures 2--4. As it was noted
above${}^3$, solutions similar to obtained one can be used only for limited time
intervals. Ultrarelativistic matter, which could dominate at a bounce, has
negligible small energy density at quasi-de Sitter stages. At the same time the
gravitating matter could be at compression stage in more realistic bouncing models,
and scalar fields could appear only at certain stage of cosmological compression.

\section{Conclusion}
As it is shown in the present paper, regular inflationary cosmological solutions for
flat, closed and open models obtained in the frame of GTG possess some interesting
physical properties, if the scale of extremely high energy densities $|\beta|^{-1}$
is much less than the Plackian one. The dynamics of cosmological models at a bounce
is determined by scalar fields and Newton's gravitational constant;
ultrarelativistic matter, which can dominate at a bounce, does not have influence on
their dynamics. It is shown, that essential part of flat, open and closed
inflationary cosmological models has bouncing character.

\section*{Acknowledgements}

I am grateful to  Alexander Garkun and  Andrey
Minkevich for  help by preparation of this paper.

\clearpage
\end{document}